\documentclass[article]{aa}
\usepackage{epsfig}
\def\cm3{cm$^{-3}$}

\def\beq{\begin{equation}}
\def\eeq{\end{equation}}

\begin{document}

\title{2D Simulations of the Line-Driven Instability in Hot-Star Winds:
\\
II. Approximations for the 2D Radiation Force}

\subtitle{}

\author{Luc Dessart\inst{1,3}
        \and
        S.P. Owocki\inst{2}
        }
\offprints{Luc Dessart,\\ \email{luc@as.arizona.edu}}

  \institute{
           Max-Planck-Institut f\"{u}r Astrophysik, Karl-Schwarzschild-Strasse,1,
           85748 Garching bei Munchen, Germany 
%           \email{luc@mpa-garching.mpg.de}
        \and
           Bartol Research Institute of the University of Delaware,
        Newark, DE 19716, USA  
%         \email{owocki@bartol.udel.edu}
        \and
           Steward Observatory, University of Arizona, 933 North Cherry Avenue, 
        Tucson, AZ 85721, USA 
%        \email{luc@as.arizona.edu}  
            }

\date{Accepted/Received}
\abstract{
We present initial attempts to include the multi-dimensional nature
of radiation transport in hydrodynamical simulations of the small-scale
structure that arises from the line-driven instability in hot-star winds.
Compared to previous 1D or 2D models that assume a purely radial
radiation force, we seek additionally to treat the lateral momentum
and transport of diffuse line-radiation, initially here within a 2D context.
A key incentive is to study the damping effect of the associated diffuse
line-drag on the dynamical properties of the flow, focusing particularly
on whether this might prevent lateral break-up of shell structures at scales
near the lateral Sobolev angle of ca. $1^{\rm o}$.
Based on 3D linear perturbation analyses that show a viscous diffusion
character for the damping at these scales, we first explore nonlinear simulations
that cast the lateral diffuse force in the simple, local form of a
parallel viscosity.
We find, however, that the resulting strong damping of lateral
velocity fluctuations only further isolates azimuthal zones,
leading again to azimuthal incoherence down to the grid scale.
To account then for the further effect of lateral mixing of radiation associated
with the {\it radial} driving, we next explore models in which the radial
force is azimuthally smoothed over a chosen scale, and thereby show that this
does indeed translate to a similar scale for the resulting density and
velocity structure.
Accounting for both the lateral line-drag and the lateral mixing in a more
self-consistent way thus requires a multi-ray computation of the radiation
transport.
As a first attempt, we explore further a method first proposed
by Owocki (1999), which uses a restricted 3-ray approach that combines
a radial ray with two oblique rays set to have an impact
parameter $p < R_{\ast}$ within the stellar core.
From numerical simulations with various grid resolutions (and $p$),
we find that,  compared to equivalent 1-ray simulations, the high-resolution
3-ray models show systematically a much higher lateral coherence.
% Because the strength of the scattered radiation field varies considerably with
% height, we anticipate that shock strength, as well as clumping factor and wind
% structure morphology may considerably vary within the flow.
This first success in obtaining a lateral coherence of wind structures
in physically consistent 2D simulations of the radiative instability
motivates future development of more general multi-ray methods that can
account for transport along directions that do not intersect the stellar core.
% are less constrained in setting the form of the radial grid to
% by the choice of radial grid and impact parameter for off-radial rays.
% a study currently under way.

}
\titlerunning{2D simulations of line-driven winds}
\maketitle

\section{Introduction}

The driving of hot-star winds by line-scattering of the star's
radiation is understood to be highly unstable to small-scale radial perturbations
(Lucy \& Solomon 1971; MacGregor, Hartmann, \& Raymond
1979; Owocki \& Rybicki 1984, 1985).
This ``Line-Driven-Instability'' (LDI) has long been supposed to be a root cause
of the extensive small-scale stochastic wind structure that has been inferred from various
kinds of observational diagnostics (see reviews by, e.g.,  
Owocki 1994; Feldmeier \& Owocki 1998; Dessart 2004a).
% (e.g., soft X-ray emission;  black troughs in saturated UV wind lines;
% nonthermal radio emission; emission line variability -- see reviews
% by, e.g.,  Owocki 1994; Feldmeier \& Owocki 1998; Dessart 2004a).
%%% EXPAND ON OBSERVATIONAL DIAGNOSTCS FOR LDI
For example, black troughs of saturated P-Cygni line profiles are
understood to arise from the net backscattering of multiple line
resonances occurring in a highly non--monotonic velocity field (Lucy 1982, 1984).
Embedded wind shocks arising from such velocity variations are
moreover thought to give rise to the soft X-ray emission often
observed from such hot stars (Chlebowski 1989; Lucy \& White 1980;
Feldmeier et al. 1997).
An even more direct diagnostic for such small-scale structure is
the explicit, small-amplitude line profile variability often
detected in high signal-to-noise (SN $\sim 100-1000$),
optical spectra obtained by ground-based monitoring of recombination
emission lines in both OB (Eversberg, L\'epine \& Moffat 1998 on
$\zeta$ Puppis) and WR stars (Robert 1992; L\'epine \& Moffat 1999).

%%%
But a key limitation in developing quantitative tests for the association
between such phenomena and the LDI is that, owing to the computational expense
of evaluating non-local integrals needed for calculating the line-force,
time-dependent dynamical simulations of the resulting nonlinear flow structure
% have either
have for many years been limited to one dimension (1D)
(Owocki, Castor, \& Rybicki 1987; Feldmeier 1995; Feldmeier et al. 1997;
Owocki \& Puls 1999; Runacres \& Owocki 2002).
% or have ignored lateral radiation transport and forces within a two-dimensional
% (2D) simulation of the hydrodynamics (Gomez \& Williams 2003;
% Dessart \& Owocki 2003, hereafter DO-03).
(See also Owocki 1999 and Gomez \& Williams 2003 for some previous
2D attempts.)
These 1D simulations show development of extensive radial variations in both
density and radial velocity, but are inherently incapable of determining
the development of any corresponding variations in the {\it lateral}
direction.
In recent years, we have focused efforts on deriving empirical
constraints on the multidimensional properties
(particularly lateral scale)
of wind structure (Dessart \& Owocki 2002a,b),
and developing approaches for multidimensional simulation of the
formation of such structure from the nonlinear evolution of the LDI
(Dessart \& Owocki 2003, hereafter DO-03).
The present effort builds on the ``2D-H+1D-R'' (denoting 2D
Hydrodynamics, but with only 1D Radiation transport)
approach of DO-03, now incorporating various
approximate attempts to account for lateral transport and
radiation forces associated with diffuse, scattered radiation,
within similar 2D hydrodynamical simulation models.

A key aspect of line-driving in hot-star winds regards the Doppler
shift associated with the wind acceleration and expansion.
Along any given direction ${\bf n}$,
this effectively desaturates the line-transfer on scales of the
directional Sobolev length
$l_{{\bf n}} \equiv v_{\rm th}/
{\bf n}\cdot \nabla ({\bf n} \cdot {\bf v})$,
where ${\bf v}$ is the velocity, and $v_{\rm th}$ is the ion thermal speed
(typically a factor few smaller than the sound speed).
% For a smooth, spherically symmetric supersonic wind expansion
% in which the flow speed $v$ is much greater than the ion thermal speed $v_{\rm th}$,
% with $v \gg v_{\rm th}$,
% In particular, for a  spherically symmetric wind expansion
% the Sobolev length along a ray that has direction cosine $\mu$ with the radial
% direction is given by $l_{\mu} = l_{0}/(1 + \sigma \mu^{2}$, with
% $\sigma \equiv d\ln v/d\ln r -1$ and $l_{0} \equiv r v_{\rm th}/v$.
In a smooth, supersonic wind with $v \gg v_{\rm th}$,
% the expansion is often nearly isotropic
% % ($\sigma \approx 0$)
% through much of the acceleration region, implying that
% for any direction
the Sobolev length
$l_{{\bf n}} \approx l_{0} \equiv r v_{\rm th}/v$ is of order
$v_{\rm th}/v \ll 1$ smaller than the characteristic wind expansion scale $r$.
The associated desaturation of the line-transfer then allows one to
write a purely {\it local} expression for the line-acceleration ${\bf g}$
in terms of the local projected velocity gradient,
${\bf n}\cdot \nabla ({\bf n} \cdot {\bf v})$,
averaged over directions to the source radiation from the stellar
core.\footnote{
A further key simplification of this Sobolev approach is that
the force associated with the diffuse, scattered component of
radiation vanishes due to fore-aft symmetries of the Sobolev escape
probability. As discussed below, the breaking of such symmetries for
structures near the Sobolev scale leads to the diffuse ``line-drag''
effect first described by Lucy (1984).}

This ``Sobolev approximation'' (Sobolev 1960) provides the basic framework
for the  standard Castor, Abbott, \& Klein (1975, CAK) model for steady,
spherically symmetric, line-driven stellar winds.
Indeed, within this generalized, vector formulation, the purely
{\it local} nature of the line-force makes even 3D
time-dependent simulations computationally feasible.
% As such, this vector Sobolev approach has been applied in numerous
% multi-dimensional models aimed at studying large-scale variations
% in line-driven winds, including those associated with, e.g.,
% rotational modulation (Cranmer \& Owocki 1995; Dessart 2004b),
% %  Dessart 2004 A&A...423..693D
% rotational distortion (Owocki, Cranmer, \& Gayley 1996;
% % Owocki, Cranmer \& Gayley 1996 ApJ...472L.115O
% Petrenz \& Puls 2000),
% % Petrenz \& Puls 2000 A&A...358..956P
% wind-wind collision (Gayley, Owocki, \& Cranmer 1997),
% % Gayley, Owocki, and Cranmer 1997 ApJ...475..786G
% and even accretion-disk outflows (Proga et al. 1999;
% % Proga, Daniel; Stone, James M.; Drew, Janet E.  1999MNRAS.310..476P
% Feldmeier \& Shlosman 1999;
% %Feldmeier, Achim; Shlosman, Isaac 1999ApJ...526..344F
% Proga \& Kallman 2004).
% % 2004ApJ...616..688P
%%% EXPANDED DISCUSSION ON OBSERVATIONAL SIGNATURES OF LARGE SCALE STRUCTURES
%%% MODELED BY MEANS OF SOBOLEV APPROXIMATION.
%%%
As such, this vector Sobolev approach has been applied in numerous
multi-dimensional models aimed at studying large-scale variations
in line-driven winds.
Associated with a rotational modulation of photospheric properties
(Cranmer \& Owocki 1995; Dessart 2004b), they reproduced some of the
key features of large scale variations observed in blueshifted
absorption troughs of UV P-Cygni profiles in O-star (Howarth \& Prinja 1989)
and Be-star (Grady, Bjorkman \& Snow 1987) winds.
Wind distortion due to stellar rotation (Owocki, Cranmer, \& Gayley
1996; Petrenz \& Puls 2000) also provides a key explanation for the observed
polar-enhanced mass loss of line-driven winds such as those of the present
day $\eta$ Car (Smith et al. 2003).
% 2003ApJ...586..432:  Smith, Nathan; Davidson, Kris; Gull, Theodore R.; Ishibashi, Kazunori;
% Hillier, D. John
The radiative-braking phenomenon (Gayley, Owocki, \& Cranmer 1997) advocates
the potential of radiation to accelerate as well as decelerate stellar winds
in massive binaries, a duality that proves essential to explain the geometry
of wind-wind collisions in massive binary systems (van der Hucht \& Williams 1995).

%%%

Unfortunately, such a {\it local} Sobolev approach cannot be used to model
structure arising from the LDI, since this occurs at scales near and
below the Sobolev length (Owocki \& Rybicki 1984, 1985).
Instead the line-force must be computed from a {\it non-local}
radiation transport that can be cast approximately in terms of
{\it integral} escape probabilities
(Owocki \& Puls 1996).
In 1D simulations, these escape probabilities can be evaluated from spatial
integrals carried out along a restricted set of near radial directions
intersecting the stellar core,
repeated for a set of $n_{x} \approx 3 v_{\infty}/v_{\rm th} \approx
1000$ frequencies that resolve the line thermal width over the full
range of velocity shifts within the wind.
This makes even 1D simulations of the LDI quite computationally expensive.

In 2D or 3D, a proper treatment of the lateral transport requires
such integration along a more complete set of oblique rays ranging from
transverse to radial in direction.
A severe complication is then that nonradial integrations from any
given grid node do not generally intersect any other grid nodes.
As such a straightforward ``long characteristic'' approach would require
repeating the integration anew for each grid node, with a complex
interpolation for the variation of the flow variables along the ray.
For even a 2D grid of $n_{r}$ radial and $n_{\phi}$ azimuthal zones,
this requires $n_{r} n_{\phi}$ integrals involving of order $n_{r} n_{x}$
operations for each of the set of $n_{ray}$ directions, giving an
overall scaling of $n_{ray} n_{x} n_{r}^{2}  n_{\phi}$ operations.
For a typical case with $n_{ray} \approx 10$, $n_{\phi} \approx 100$,
and $n_{x} \approx n_{r} \approx 1000$, this implies of order
$10^{12}$ operations to evaluate the radiative force {\it at each time
step} of a simulation model!

Such timing might be reduced somewhat by a ``short characteristic''
approach that builds up the local escape probabilities based on the
evaluation in neighboring zones
(e.g., van Noort, Hubeny, \& Lanz 2002).
% vanÊNoort,ÊMichiel; Hubeny,ÊIvan; Lanz,ÊThierry 2002 ApJ...568.1066V
But before attempting to develop such a complex and costly general
method, we explore here some more tractable, approximate treatments
for the lateral radiation transport and associated force, aimed at gaining
some initial insights into the key dynamical effects within 2D models.

In particular, a key issue in such instability simulations is
taking proper account of the ``line-drag'' effect of the diffuse,
scattered radiation (Lucy 1984; Owocki \& Rybicki 1985).
In 1D simulations, the associated reduction in the net growth rate of
the strong radial instability can be modeled via a ``Smooth Source
Function'' (SSF; Owocki 1991, Owocki \& Puls 1996) method.
This ignores any variations in the scattering source function,
but accounts, through the radial integrations for the non-local escape
probability, for small-scale, fore-aft asymmetries that give rise to this
diffuse drag
% for velocity variations at scales near and below the Sobolev length
(Owocki \& Rybicki 1985).
This nearly stabilizes the flow near the wind base, but in the outer
wind there is still a  strong  net  radial instability from driving by the
direct radiation from the stellar core.

For {\it lateral} directions not intersecting the stellar core,
the 3D linear stability analysis by
Rybicki, Owocki, \& Castor (1990, hereafter ROC)
shows that this diffuse line-drag leads to a strong net {\it damping}
of velocity variations at scales near and below the
lateral Sobolev length $l_{0} = r v_{\rm th}/v$.
In DO-03 we speculated that such damping might inhibit the lateral
overturning of Rayleigh-Taylor and thin-shell instabilities that break
up radial shell structures, and so might lead to an overall lateral coherence
at an associated angular scale
$\Delta \phi_{0} \approx l_{0}/r = v_{\rm th}/v \approx 0.01$~rad
$\approx 1^{o}$.
This ``Sobolev angle'' is somewhat smaller than, but comparable to,
the typical angle scale $\sim 3^{o}$ inferred for wind structure from
analysis of line-profile variations in emission lines from Wolf-Rayet stars
(Dessart \& Owocki 2002a,b).

% Based on a recasting (Appendix A) of the ROC linear stability analysis
% to show that the lateral line-drag takes the form of a viscous damping at
% scales near the Sobolev length,

% A central goal of the present paper here is
To explore this and other
effects that might influence the lateral scale of structure,
the present paper carries out
2D instability simulations that account approximately for lateral
transfer effects, including the diffuse line drag.
Our initial approach  (Sect. 2) uses a simple, local
{\it parallel viscosity} formulation for the azimuthal diffuse
line-force;
as shown in Appendix A, the 3D linear stability analysis by ROC
implies such a viscous scaling for the line force arising from azimuthal
velocity perturbations on a scale near and above the lateral Sobolev
length $l_{0}$.
Results from our nonlinear simulations show that this lateral viscosity
can indeed strongly damp azimuthal velocity
variations, but does not, by itself, lead to a lateral coherence above
the azimuthal grid scale.

Arguing then that such coherence might instead arise from lateral
mixing of radiation associated with the {\it radial} driving, we next
explore (Sect. 3) models with an azimuthal smoothing of the radial
line-force, showing that this does lead to a comparable lateral
smoothing of the resulting flow structure.
To account more consistently for both effects, we finally examine fully
non-local formulations of both the radial and azimuthal force obtained
using a restricted, grid-aligned, ``3-ray SSF method'' first introduced
by Owocki (1999).
Initial results do show an extended lateral coherence of
instability-generated structure, but inherent limitations in
the ray coverage and outer radial grid resolution in the method leave
uncertain the broad applicability of these 3-ray simulations.
We conclude (Sect. 5) with a brief summary of results and their
implications for future development of generalized multi-ray methods
to account for multi-dimensional transport in simulations of
structure arising from the LDI.

 \begin{figure*}
 \vspace{10cm}
 \hspace{1cm}
% \special{psfile=d_v01_v02_v03.eps}
% \special{psfile=2778f1.eps}
 \caption{
Grayscale representation of the density contrast (normalized to
the initial relaxed CAK model obtained with identical wind and stellar parameters) for
a sequence of models with distinct viscosity amplitude $\alpha_{\rm vis}$ (see Sect. 1 and Appendix),
increasing clockwise over the range 0.01, 0.1 and 1.
For readability, each computed wedge is duplicated and stacked four times in azimuth,
using the periodic boundary conditions employed in that direction.
}
\end{figure*}

\begin{figure*}
 \vspace{4.5cm}
% \special{psfile=v01_stat.eps}
 \includegraphics{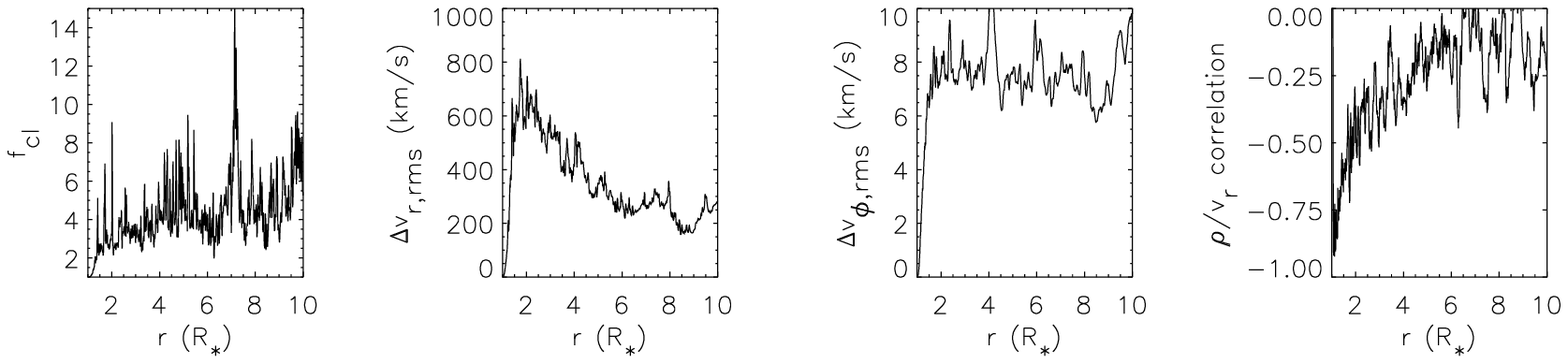}
 \vspace{4.5cm}
% \special{psfile=v02_stat.eps}
 \includegraphics{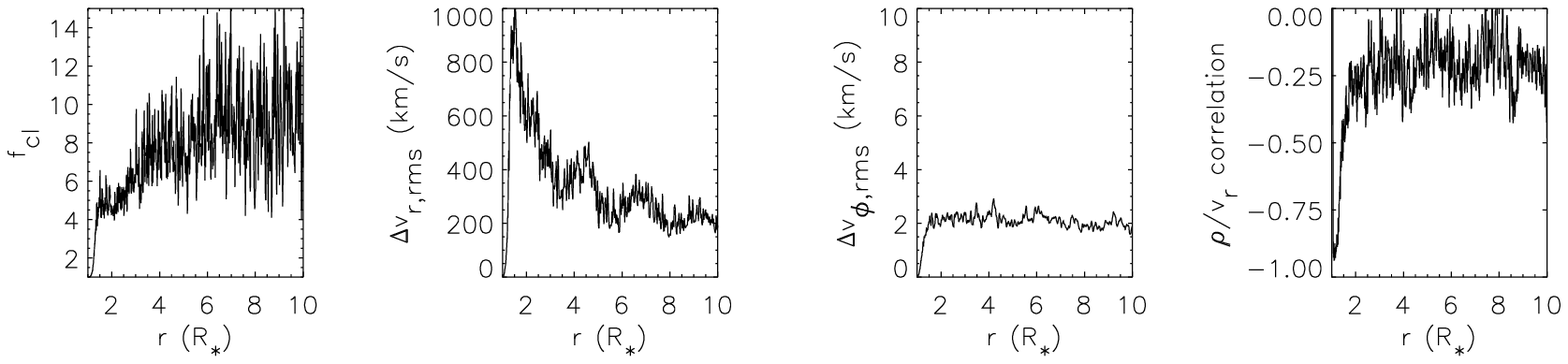}
 \vspace{4.5cm}
% \special{psfile=v03_stat.eps}
 \includegraphics{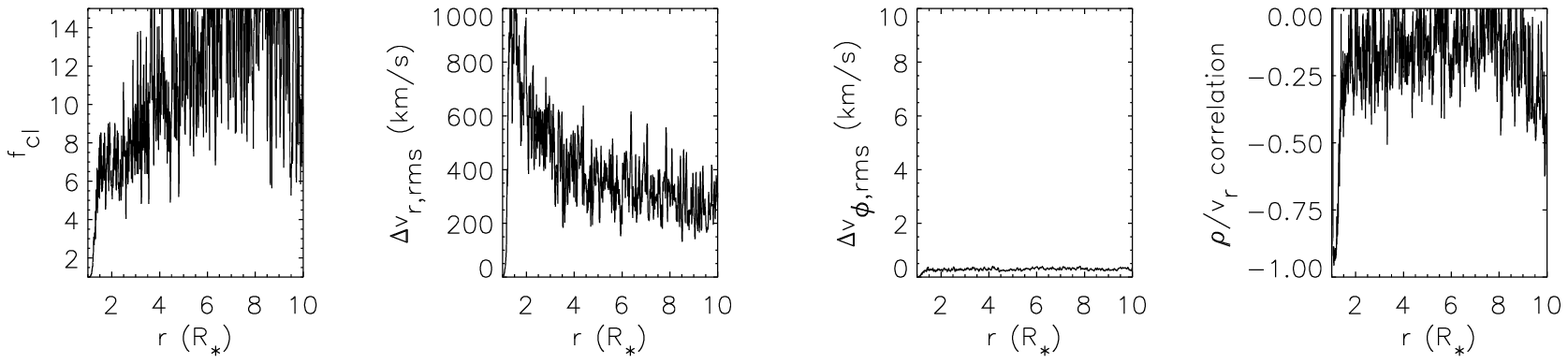}
 \caption
{
{\bf Top row} Radial variation of time- and angle-averages that characterize the
nature of flow structure for the case with $\alpha_{\rm vis}=0.01$.
{\bf a)} clumping factor $f_{\rm cl}$, {\bf b)} radial and {\bf c)} azimuthal velocity dispersion,
and {\bf d)} velocity-density correlation coefficient (Runacres \& Owocki 2002).
{\bf Middle row} Same as top for a model with $\alpha_{\rm vis}=0.1$.
{\bf Bottom row}  Same as top for a model with $\alpha_{\rm vis}=  1$.
For ease of comparison, corresponding quantities are plotted over the same ordinate range.
}
\end{figure*}

\section{Radiative Viscosity Model for the Lateral, Diffuse Force}

\begin{figure*}
 \vspace{6cm}
% \special{psfile=v01_force_comp.eps}
 \includegraphics{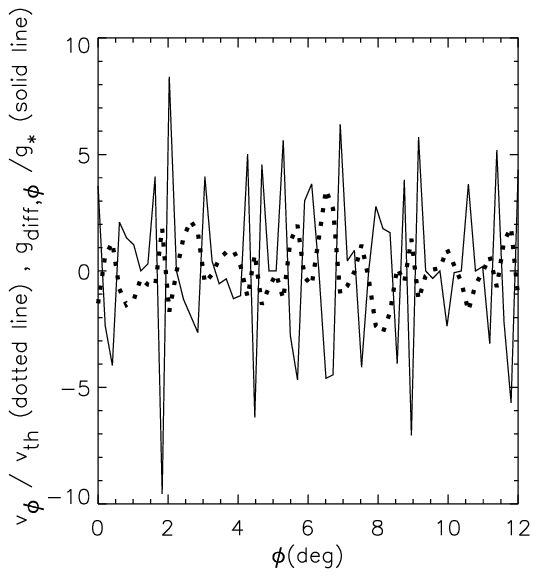}
 \hspace{6cm}
% \special{psfile=v02_force_comp.eps}
 \includegraphics{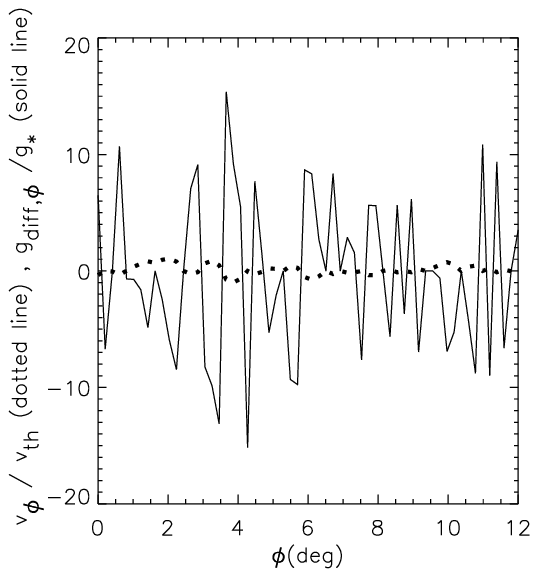}
 \hspace{6cm}
% \special{psfile=v03_force_comp.eps}
 \includegraphics{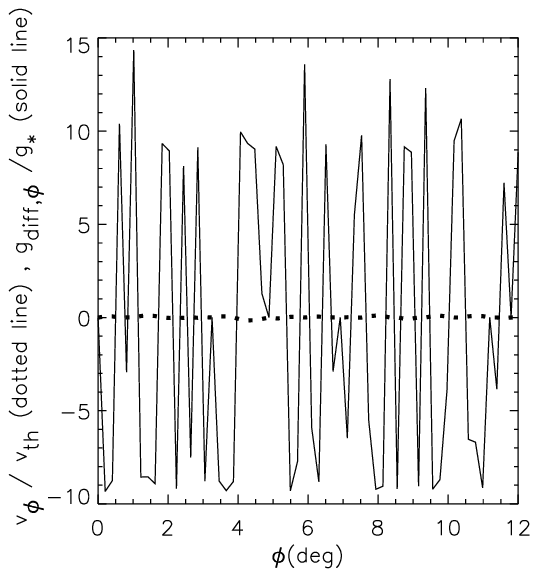}
 \caption
{
{\bf Left} Model with $\alpha_{\rm vis}=0.01$.
Azimuthal variation of the normalized viscous diffuse force
$g_{\rm diff,\phi}(r,\phi) / g_{\ast}(r)$, at $r=6 R_{\ast}$,
and corresponding to the time-snapshot of Fig. 1.
The lateral velocity, normalized to the thermal speed, is over-plotted (dotted line).
{\bf Middle} Same as left for a model with $\alpha_{\rm vis}=0.1$.
{\bf Right}  Same as left for a model with $\alpha_{\rm vis}=  1$.
}
\end{figure*}

As first derived in the 3D linear stability analysis by ROC,
a key multidimensional effect of the diffuse radiation is the tendency
to strongly damp lateral velocity perturbations.
Appendix A shows that this can be cast in a general analytic form
(Eq. \ref{A3}) that reduces to a simple viscous diffusion
(Eq. \ref{A5}) for variations  on scales near or above the (quite small)
Sobolev length, $l_{0} \approx r v_{\rm th}/v \approx 0.01 r$.
To now account for the dynamical effects of diffuse lateral transport
in {\it nonlinear} simulations of wind structure,
let us here mimic this scaling from the linear perturbation analysis,
and so assume that the azimuthal component of the diffuse line-force
can be approximated as a standard parallel viscosity term,
\beq
g_{\rm diff,\phi}({\bf r}) =
\alpha_{\rm vis} (v_{\rm th} r ) ~
{ \partial^2 v_\phi({\bf r}) \over r^{2} \partial\phi^2}  \, .
\label{eq1}
\eeq
Here $\alpha_{\rm vis}$ is a dimensionless parameter to control the
overall strength of the viscous dissipation in terms of the assumed dimensional
scale $v_{\rm th} r$ for the kinematic viscosity.
This is indeed the scaling derived from the linear
perturbation analysis (cf. Eq. \ref{A5}), with $\alpha_{vis} \approx
s \alpha/6 $ a fraction of order a few tenths.
% predicted to be of order unity.

More generally, this approach is also consistent with the notion that any
moderate- to large-scale net asymmetry in
the diffuse radiation field should be set by departures from the
Sobolev limit defined by $v_{\rm th}/v \rightarrow 0$.
% which applies in smooth, highly supersonic outflows.
In particular, following the standard Sobolev approximation to next
highest order in this small parameter leads to corrections that scale
with $v_{\rm th}$ times the gradient of the Sobolev optical
depth $\tau \sim \kappa \rho v_{\rm th}/v'$, where $v'$ is the
velocity gradient along the direction of interest.
For lateral components in which the gradient operator $\nabla \sim
\partial /\partial \phi$, the combined derivatives lead to terms that scale as
$v_{\rm th} \partial^{2} v_{\phi}/\partial \phi^{2}$, as in Eq. (\ref{eq1}).

Using this formulation, we have performed simulations analogous to the 2D-H+1D-R
models described  in DO-03, but now including, for negligible additional CPU costs,
this viscous approximation (\ref{eq1}) for the azimuthal diffuse radiation force,
assuming various values of the dimensionless coefficient $\alpha_{\rm vis}$.
The radial and azimuthal grids are as defined in
DO-03, with $n_{\phi} = 60$ azimuthal zones of
constant angular size $\delta \phi = 0.2^{\rm o}$, thus extending over a wedge
of $\Delta \phi = n_{\phi}\times \delta \phi =12^{\rm o}$, with periodic
boundary conditions in azimuth.

As in the 2D-H+1D-R case, the simulations begin from a smooth, relaxed CAK model,
from which there is initial formation of laterally coherent shell structures
that arise from the strong  radial instability of the line driving.
Over time, these shells again break up from thin-shell and
Rayleigh-Taylor instabilities, but with a final asymptotic form that
is quite different depending on the assumed value of $\alpha_{\rm  vis}$.
Figure 1 compares a representative snapshot of this asymptotic density structure
(normalized by the density in the smooth CAK initial condition) for
three models, divided into three display wedges of $60^{\rm o}$
(representing 5 repetitions of the $12^{\rm o}$ model computational wedges),
corresponding in clockwise order to models with $\alpha_{\rm vis} =$~0.01, 0.1,
and 1.
Comparison with the top panel of Fig. 1 in DO-03 shows that the
structure in the low viscosity case $\alpha_{\rm vis} = 0.01$
(leftmost wedge of Fig. 1 here)
is quite similar to that obtained in this previous model without any
lateral radiation forces;
but the more viscous cases with $\alpha_{\rm vis} = 0.1$ and
$\alpha_{\rm vis} = 1$ (middle and right wedges of Fig. 1) are quite
different, with progressively less radial elongation in
filamentary structures.

To provide a more quantitative comparison, Fig. 2 shows
representative statistical properties derived from time- and angle-averages of
the density and velocity fields.
For the case where $\alpha_{\rm vis} = 0.01$ (left wedge in Fig. 1 and top row
in Fig. 2),
the amplitude of azimuthal velocity variations is slightly reduced,
but otherwise
the clumping factor $f_{\rm cl}$,
the radial velocity dispersion $\Delta v_{r, \rm rms}$,
and the velocity--density correlation function
are all similar to those given in DO-03
with no lateral viscous term ($\alpha_{\rm vis} = 0$).

By contrast, for the stronger viscosity cases
$\alpha_{\rm vis} = 0.1$  and $\alpha_{\rm vis} = 1$
(middle and lower panels of Fig. 2),
we see that the azimuthal velocity is markedly reduced,
with associated changes also in the other statistical parameters.
The inhibition of lateral motion in effect tends to isolate further
each of the azimuthal coordinates.
In this situation, when a shock occurs, material from below ramming
into the dense structure is prevented from circumventing it by the
stronger radiative viscosity in the lateral direction.
% In this situation, when  faster flow rams into the slower, denser
% material the lateral viscosity now inhibits the
% from circumventing it
This behavior becomes even more pronounced in the highest  viscosity
case ($\alpha_{\rm vis}=1$), for which the dense structures resemble both radially and azimuthally confined
clumps, i.e. dots in 2D.
As shown in the bottom row of Fig. 2, the azimuthal velocity dispersion is then reduced by a factor of
20--30  compared to the first case.
This corresponds to about a tenth of the sound speed, i.e. three orders of magnitude less than
the wind flow speed.
We see also that the clumping and radial velocity dispersion increase significantly with $\alpha_{\rm vis}$,
converging for high $\alpha_{\rm vis}$ to values similar to those found in purely 1D non-Sobolev simulations.
Then, the lateral communication becomes so inhibited that each direction is essentially sheltered from
its neighbors, and the 2D computed grid merely looks like a series of
1D non-Sobolev simulations stacked together in azimuth.

For a characteristic wind radius $r=6 R_{\ast}$,
Fig. 3 compares the lateral variation of the
azimuthal velocity and lateral radiation force for these three
values of $\alpha_{\rm vis}$.
Note that the magnitude of the viscous forces are comparable (solid
curves), even though the associated azimuthal velocity (dotted curves)
is much smaller for the higher viscous coefficient.
Note also that the sign of the viscous force depends on the concavity
of the velocity variation.

% Such a radiative viscosity has therefore the desirable property of limiting the lateral
% communication between adjacent directions, and therefore possesses a promising ability
% to prevent the growth of thin-shell or Rayleigh-Taylor (RT) instabilities.
% However, by de-correlating different azimuthal directions, fluctuations in the outer
% wind act as perturbation seeds at the wind base but over azimuthally confined regions.
% Such a de-correlated backscattered radiation field eventually leads to the formation of
% radiative-instability generated structures whose lateral extent is limited to the azimuthal
% resolution scale, i.e. completely set by (arbitrary) numerical choices.

A key result here is that the lateral diffuse radiation has a net
effect that, in some sense, is opposite of what was anticipated, e.g.,
in DO-03.
%LUC: HAVE OTHERS WEIGHED IN ON THIS?? Feldmeier 1998 on 2nd order Sobolev
% discusses the instability properties for core rays only.
In particular, because the line-drag of diffuse radiation strongly damps
lateral velocity variations on scales smaller than the lateral Sobolev length
$l_{0} \approx r v_{\rm th} /v$, DO-03 speculated that this could inhibit
the operation of Rayleigh-Taylor or thin-shell instabilities on this
scale, and thus lead to a finite azimuthal coherence at the associated
angle scale $\Delta \phi_{\rm Sob} \approx l_{0}/r = v_{\rm th}/v$.
However, the above results show that such diffuse radiation damping, as
modeled here in terms of a lateral viscosity, tends instead to limit
further the lateral communication between neighboring azimuthal zones,
and thus lead to an even greater level of lateral {\it incoherence}.

Since lateral line-drag does not limit structure to a finite
azimuthal scale, we thus next consider how this might instead arise
from the angle averaging of the backscattered radiation,
which we next model in terms of an azimuthal averaging
of the radial driving force.

% Unlike DO-03, who speculated that the thin-shell or RT instability in competition with the
% lateral line-drag effect would be the key ingredients conditioning the lateral scale of wind
% structures,  we now believe that the lateral coherence of the radiation field might also
% play an essential role.
% In reality, each wind point feels an {\it angle-averaged} backscattered radiation
% field, inducing both reduced and more coherent fluctuations of the line-driven fluid.
% Hence, we investigate in the next section the impact of an enforced lateral coherence
% in the radiation field on the 2D properties of a line-driven wind.

% \section{Artificial lateral coherence for diffuse/direct radiation force terms}

\section{Azimuthal Averaging of the Radial Radiation Force}

\begin{figure*}
 \vspace{16cm}
 \hspace{1cm}
% \special{psfile=d_f1_f2_f2p5_f3_f4_new.eps}
% \special{psfile=2778f4.eps}
 \vspace{-1cm}
 \caption
{
Grayscale images of the density contrast normalized to
an relaxed CAK model with identical wind and stellar parameters.
At the top (bottom), models are shown only out to 10 (3) R$_{\ast}$.
Each panel is composed of five different models (duplicated and stacked two times
laterally to improve the visibility), ordered clockwise with
increasing $\sigma_{\phi}$, i.e. 0, 0.1, 0.5, 1 and $1.5^{\rm o}$.
}
\end{figure*}

A key limitation of the above lateral viscosity approach is that it
still does not account for ways in which the lateral radiation
transport might alter the {\it radial} component of the line-driving
force.
Indeed, since the radial driving at each azimuth is solved
independently by strictly radial integrations of the line optical depth,
there arises, through the diffuse component of the radial force,
a backscattering feedback between the outer and inner wind that is
effectively isolated and independent for each azimuthal zone.
As such, both the previous 2D-H+1D-R and the present lateral
viscosity models tend to develop structure that is independent
for each azimuthal zone, implying a lateral incoherence down to the
azimuthal grid scale.

More realistically, such radial backscattering should incorporate
a nonzero level of lateral averaging, associated with
transport along oblique rays that couple zones of different radius
{\it and} azimuth.
Unfortunately, as discussed in the Introduction, there are severe
computational challenges to accounting for such oblique transport
through direct integration along non-radial rays.

Thus, as a first approximate attempt to explore such effects, we have
carried out here simulations in which the radial force is simply
averaged over azimuth using a Gaussian filter with a tunable
``smoothing angle'' $\sigma_{\phi}$.
At a given azimuthal coordinate $\phi_{j}$ and radial
coordinate $r$, the smoothed radial force takes the  form,
\beq
g_{\rm r,smooth} (r,\phi_j) =
{
\Sigma_{k} g_{\rm r} (r,\phi_k)
e^{-[(\phi_j - \phi_k)/\sigma_{\phi}]^2}
\over
\Sigma_{k}  e^{-[(\phi_j - \phi_k)/\sigma_{\phi}]^2}
}\, .
\label{eq2}
\eeq

% We ran a sequence of models for values of $\Delta \phi$ of 0, 0.1,
% 0.5 and 1 and $1.5^{\rm o}$, for which we show the normalized density in
% Fig. 4 (each simulated wedge has been duplicated and stacked two times to improve the
% visibility of the figure).

Figure 4 compares the normalized density structure for simulations
with $\sigma_{\phi}$ of 0, 0.1, 0.5, 1 and $1.5^{\rm o}$,
represented respectively by the clockwise sequence of wedges of
width of $36^{\rm o}$, each of which corresponding to 3 azimuthal periods of
$\Delta \phi = 12^{\rm o}$ for each of the 5 model cases.
The upper panels show results for the full simulated grid (out to 10$R_{\ast}$),
while the lower panels  zoom in on the region between the photosphere
and 3$R_{\ast}$.
Note that cases with a smoothing angle $\sigma_{\phi}$ less than the lateral grid
resolution  $\delta \phi = 0.2^{\rm o}$ show little departure from standard
2D-H+1D-R simulations.
However, as $\sigma_{\phi}$ increases above $\delta \phi$, the coherence of wind
structures becomes more pronounced.
Indeed, for a smoothing angle of $\sigma_{\phi} = 3^{\rm o}$,
or twice the maximum value shown in Fig. 4, we have found the lateral coherence
extends across the full azimuthal wedge of $12^{\rm o}$, so that the model
effectively recovers the radially confined concentric shells
%seen in purely 1D instability simulations.
corresponding to purely 1D (spherically-symmetric) instability
simulations.
%
% For a smoothing angle $\Delta \phi = 3^{\rm o}$, radiative-instability generated
% structured have a lateral coherence that stretches throughout the grid,
% which covers 12^{\rm o}$ (not shown
% since this looks like radially confined concentric shells).
Overall, this experimentation, although artificial, highlights the potential role
of azimuthal averaging of radial driving in setting the lateral coherence of
radiatively driven wind structures.

\begin{figure}
 \vspace{10cm}
 \hspace{-1.25cm}
% \special{psfile=grid_nr74_dphi0.9deg_np100.eps}
 \includegraphics{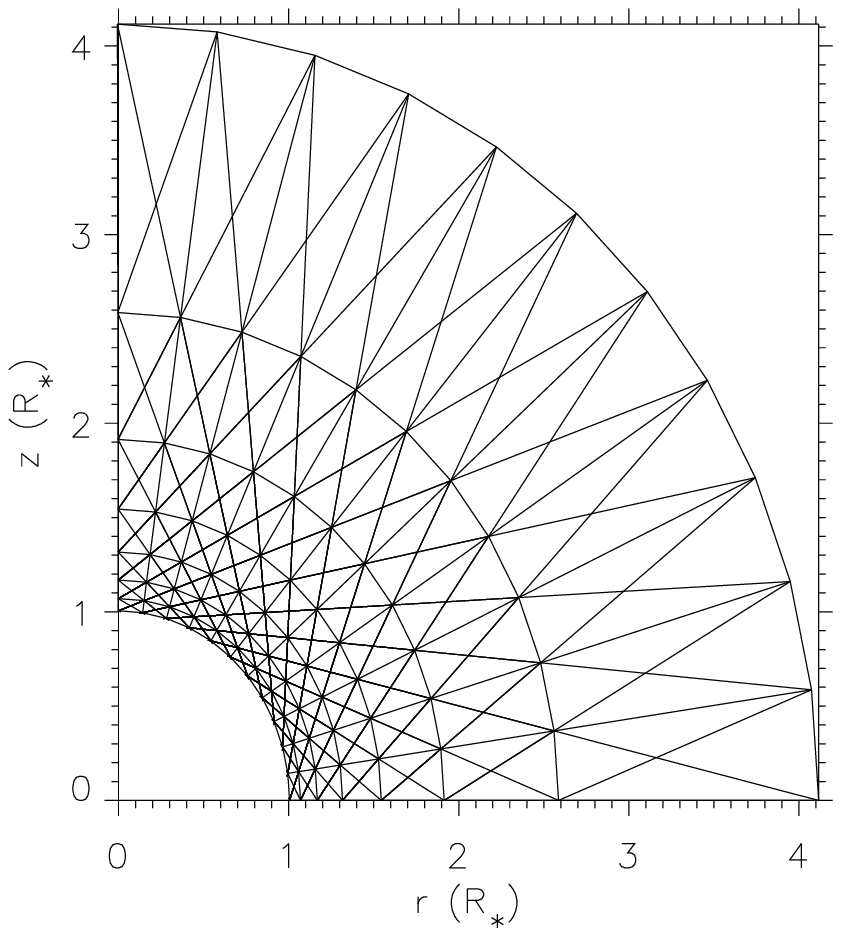}
 \caption
{
Illustration of 3-rays with $ p/R_{\ast} = 0, \, \pm \sqrt{0.9}$ spaced
azimuthally every  $\delta \phi =0.9^{\rm o}$, together with the radial grid
spacing (circular arcs) that aligns the rays to intersect multiple $(r,\phi)$ grid
nodes; this allows very efficient evaluation of the ray escape integrals
needed for computation of the SSF line-force in 2D wind models.
For clarity, only every ninth azimuthal or radial zone is shown.
%
% Typical grid morphology imposed by the 3-ray formalism discussed in Sect.~4. The
% parameters are $\delta \phi = 0.9^{\rm o}$, $p/R_{\ast} = \sqrt{0.9}$
% and $nr=74$ (rays are plotted only every nine azimuthal and radial grid points).
% We limit the outer radius to ca. 4$R_{\ast}$ to allow a good visibility
% of the inner region.
% The advantage of this grid sizing is that each ray goes through
% a grid point at each radial shell crossing, therefore not requiring any
% interpolation.
% However, notice how fast the radial increment increases with height, resolving
% accurately the wind base but only coarsely the outer wind.
}
\end{figure}

\section{Simulations Using a 3-Ray Method for the Nonlocal 2D Line-Force}

\subsection{Method Formulation}

% The above models attempt to account for the effects of lateral transport
% on the radiative force through either a local viscosity or a simple
% smoothing, while using a more robust escape integral formulation for
% the radial force.

While the above models use a quite intricate Smooth-Source-Function
(SSF) escape-integral  formulation for the {\it radial} force along
each azimuth, their account of {\it lateral} transport effects on the
radiative force is only phenomenological, through either a local
viscosity or a simple azimuthal smoothing.
A more consistent approach would carry out similar escape
integrals along an appropriate set of oblique rays spanning the range
from radial to lateral directions.
Unfortunately, as already noted in Sect. 1, such a calculation presents severe
computational challenges, stemming largely from the general
misalignment of these rays with the nodes of the
computational grid.

However, as first introduced by Owocki (1999), there is a specialized
spatial  grid that can allow a tractable, 3-ray, non-local formulation of
both the radial and lateral components of the line-force.
% As illustrated in Fig. 5,
% through each grid point there are two oblique rays set
% to have a fixed impact parameter $p$ to the origin on opposite sides
% of the single radial ray.
% A key trick is here to chose the radial grid spacing in such a way
% that, for a given choice of the fixed azimuthal grid size
% $\delta \phi$, the three rays intersect at each grid node.
% From simple trigonometry, one finds the radius coordinates is set by
% \beq
%  r_1 = R_{\ast} ~~ ; ~~ r_i = \frac{p}{\sin (\arcsin(p/r_{i-1})- \delta \phi)}  \, , i=2,n_r
% \eeq
At any given grid point, the non-local escape probabilities are
evaluated along one radial ray, plus two nonradial rays on opposite sides of the
radial direction, set always to have a fixed impact parameter
$p< R_{\ast}$ toward the stellar core.
A key trick is to {\it choose} the radial spacing so that each nonradial ray intersecting
a grid point with indices $\{i,j\}$ will also intersect other grid points $\{i\pm n,j\pm n\}$,
for integer $n \ge 1$.
This avoids the need to carry out a conceptually complex and computationally
costly interpolation between a $(p,z)$ ray grid for the radiation transport,
and the $(r,\phi)$ grid for the hydrodynamics.
For uniform azimuthal spacing $\delta \phi$, such
ray alignment occurs for radial grids satisfying
\beq
    r_i = { p \over \cos{\left [ i \delta \phi + \arccos (p/R_{\ast})
\right ]} } \, .
\label{eq3}
\eeq
Figure 5 illustrates the grid and ray alignment for
$\delta \phi = 0.9^{\rm o}$,
$p/R_{\ast} = \sqrt{0.9}$ and $n_r=74$.
Through spatial integration along the 3 such rays for each of the $n_{\phi}$ azimuthal
zones, one obtains a ``6-stream'' description (i.e. in 2 directions along
each of the 3 rays) for the required non-local escape probabilities
from each of the $n_{r} \times n_{\phi}$ grid nodes.

Because each ray integration provides the escape probabilities along
ca. $n_{r}$ grid nodes, the overall timing is now {\it linear}
(not quadratic, see Sect. 1) in $n_{r}$, scaling overall as
$3 n_{\phi} n_{r} n_{x} \approx 10^{8}$ operations, or ca. a factor $10^{4}$
savings over the ca. $10^{12}$ operations estimated for a
``brute-force'', long characteristic computation in a non-aligned grid
(cf. Sect. 1).

The difference in escape along the two nonradial rays provides
a rough treatment of the lateral radiation transport.
Because these rays are restricted to always impact within the stellar core
radius ($p<R_{\ast}$), they are best suited for approximating the {\it direct} component
of the  line-force; but in the crucial wind-acceleration region near the star, they
have a substantial azimuthal component, and so also provide a rough approximation
of the azimuthal part of the {\it diffuse} line-force, including, for example,
the important lateral ``line-drag'' effect that is predicted to strongly damp
small-scale azimuthal velocity variations (ROC).
As the rays become increasingly radial at larger radii, this capacity to
approximate the lateral, diffuse radiation is lost, but the 3 rays still provide
a quite accurate representation of the finite-disk form for the direct line-force.

A more serious limitation arises from the severe loss of radial resolution at large
radii, as demonstrated by the radial/azimuthal grid aspect ratio,
\beq
{\delta r_i \over r_i \delta \phi } \approx \sqrt{(r_i/p)^2-1} ,
\label{eq4}
\eeq
which increases as $r_i/p$ at large radii.
This means that small-scale radial structure can be relatively well resolved
in the inner wind, but then becomes increasingly more damped by grid averaging in
the outer wind.

In our general scenario of self-excited instability, the nonlinear structure
at larger radii plays a key role in seeding perturbations in the
inner wind, through the backscattering feedback of the diffuse
line-force.
Thus it is important to minimize as much as possible this level of outer
grid damping, by choosing as fine as possible radial grid, which in
the 3-ray formalism here requires using as small as possible
azimuthal grid size $\delta \phi$.
In the simulations described below, we thus choose
% very small angular resolution scales
$\delta \phi \approx 10^{-3} {\rm rad} \approx
0.06^{\rm o}$  (see Table 1)
that are even finer than in previous models.
% (for which the principal requirement was to resolve the {\it lateral} Sobolev scale.)
For a characteristic wind Sobolev length
$l_{0} \approx r  v_{\rm  th}/v \approx 0.01 r$, this implies the radial grid size
$\delta r$ is sufficient to resolve unstable structure near this scale out to a
radius $r\approx p/100 \delta \phi \approx 10 p \approx 7 R_{\ast}$.

% From the frequency-dependent optical depth along the 3 rays, one can obtain the
% corresponding integral escape probabilities in the outward and inward directions,
% and then from the flux moment of these, compute the  direct and diffuse
% contributions to the line-force. (See eqs. 65 and 67 of Owocki and Puls 1996.)
% In addition to the radial line-force that drives the wind outflow, there is, in
% general, a nonzero {\it azimuthal} force component as well.

% While the resolution near the wind base is high and comparable in both radial
% and azimuthal directions, the former becomes poorer with increasing height.
% To resolve the radial and lateral Sobolev length at mid-wind heights,
% one must employ $\delta \phi \la 10^{-3}$rad (see Sect. 1),
% which typically corresponds to a full azimuthal extent of ca. $3^{\rm o}$, thus
% only marginally greater than the inferred lateral scale of wind structures.
% Hence, the numerical constraints imposed by the grid set-up are strong.
% Here, we demonstrate that it is nonetheless useful to perform such 3-ray simulations
% to gain a better understanding of the behavior and the dynamical impact of the radiation
% force in the lateral direction.

\begin{table}\caption[]{Grid parameters used in the 3-ray SSF models. All simulations
use 60 azimuthal zones and cover the first 10 stellar radii. We also give the
extrema of the radial and azimuthal resolutions, to compare with the Sobolev
length at mid wind-heights of ca. 0.01 R$_{\ast}$ along the three rays.}
\label{tab1}
\begin{tabular}{lcccccc}
\hline
Model &  \multicolumn{2}{c}{$\delta r / R_{\ast}$} &$n_r$& $p/R_{\ast}$ &
$\delta \phi$ (rad)\\
      &      min.  & max. &           &               &                       \\
\hline
% r02
A        & 2.2 10$^{-4}$  & 0.074 &  1737   & $\sqrt{0.9}$    & 6.65 10$^{-4}$  \\
% r01
B        & 6.7 10$^{-4}$  & 0.101 &  1076   & $\sqrt{0.5}$    & 6.65 10$^{-4}$  \\
% r04
C        & 4.5 10$^{-4}$  & 0.155 &  869    & $\sqrt{0.9}$    & 1.33 10$^{-3}$  \\
% r03
D        & 1.3 10$^{-3}$  & 0.220 &  539    & $\sqrt{0.5}$    & 1.33 10$^{-3}$  \\
\hline
\end{tabular}
\end{table}

%cccccccccccccccccccccccccccccccccccccccccccccccccccccccccccccccccccc
%c  xpp=sqrt(0.5), np = 60, nr = 539, Rmax= 10., delphi= 1.33e-3,
%c                 x1min=1.3224e12, x1max=set by choice of ,
%c                 x3min= 1.5309, x3max=1.6107, phirange=4.57deg
%c                 in init.dat: idump = 5
%c
%c  xpp=sqrt(0.5), np = 60, nr = 1076, Rmax= 10., delphi= 6.65e-4,
%c                 x1min=1.3224e12, x1max=set by choice of ,
%c                 x3min= 1.5508, x3max=1.5907, phirange=2.26deg
%c                 in init.dat: idump = 5
%ccccccccccccccccccccccccccccccccccccccccccccccccccccccccccccccccccccc
%c
%c  xpp=sqrt(0.9), np = 60, nr = 869, Rmax= 10., delphi= 1.33e-3,
%c                 x1min=1.3224e12, x1max=set by choice of ,
%c                 x3min= 1.5309, x3max=1.6107, phirange=4.57deg
%c                 in init.dat: idump = 5
%c
%c  xpp=sqrt(0.9), np = 60, nr = 1737, Rmax= 10., delphi= 6.65e-4,
%c                 x1min=1.3224e12, x1max=set by choice of ,
%c                 x3min= 1.5508, x3max=1.5907, phirange=2.26deg
%c                 in init.dat: idump = 10
%ccccccccccccccccccccccccccccccccccccccccccccccccccccccccccccccccccccc

% \subsection{The diffuse force for ideal situations: the long and short wavelength limit}
\subsection{Linear Perturbation Test for 3-Ray Azimuthal Diffuse Force}

Before applying this 3-ray model in simulations of the nonlinear
evolution of wind structure, it is instructive to examine its response
to linear, test perturbations applied in a smooth, CAK initial
background model.
For this, let us compute the diffuse force for the simple case of a
small amplitude lateral velocity perturbation $v_{\phi}$
that has a Gaussian variation in both
radius and azimuth, centered on a grid coordinate ($ r_{\rm c},\phi_{\rm c}$),
with the corresponding widths $\Delta r_{\rm c}$
and
$\Delta \phi_{\rm c}$,
\beq
v_{\phi}(r,\phi) = v_{\rm th}
\exp \left [ - \left ( \frac{ r-r_{\rm c}}{\Delta r_c} \right )^2
 - \left (\frac{ \phi-\phi_{\rm c}}{\Delta \phi_{\rm c}} \right )^2 \right]
\, .
\label{eq5}
\eeq
Specifically, let us choose a representative radius
$r_{\rm c}=1.5 R_{\ast}$, with $\phi_{\rm c}$ set at the mid-angle of the
azimuthal grid,
and then consider two cases distinguished by the relative scale of
the assumed widths.
For both cases, the 3-ray grid parameters $p$ and $\delta \phi$ are taken from
the set B in Table 1.

The first case assumes a very small scale for both the radial and azimuthal
widths,  $\Delta r_c = 0.0005R_{\ast}$ and $\Delta \phi_0=0.0001$~rad,
smaller in fact than the corresponding grid sizes, so that in effect the
perturbation is essentially confined to the single grid point at
($r_{\rm c}, \phi_{\rm c}$).
Since both scales are thus also much smaller than the
characteristic Sobolev length
$l_{0} \approx r v_{\rm th}/v \approx 0.01 r$,
the response should follow the direct damping scaling given by Eq. (\ref{A4})
of the linear perturbation analysis in Appendix A.
The lower panel of Fig. 6 shows just this type of scaling for this
small-width case of the 3-ray numerical model,
with the resulting azimuthal force (solid curve) varying with the
{\it negative} of the azimuthal velocity perturbation (dashed curve).

The second case assumes a very large radial extent
$\Delta r_{\rm c} = 0.2R_{\ast}$,
and an azimuthal extent $\Delta \phi_{\rm c}=0.008$~rad that
corresponds roughly with the characteristic Sobolev angle
$l_{0}/r \approx 0.01$.
With such comparatively large scales, the linear perturbation analysis
in Appendix A now predicts the response to have the diffusive scaling (Eq. \ref{A5}), which
implies a force that varies as the {\it second derivative} of the applied
Gaussian velocity perturbation.
The upper panel of Fig. 6 shows that the 3-ray response in this
larger-width case does indeed have just this type of scaling.

Overall, these tests demonstrate that, despite the inherently restricted
nature of the angle quadrature, the 3-ray method does properly reproduce the
appropriate scalings for the azimuthal component of the diffuse line-force.
We thus now proceed to apply this method in simulations of the
nonlinear evolution of the 2D flow structure.

\begin{figure}[!ht]
 \vspace{18cm}
 \hspace{0.5cm}
% \special{psfile=short_long_limit_3ray.eps}
 \includegraphics{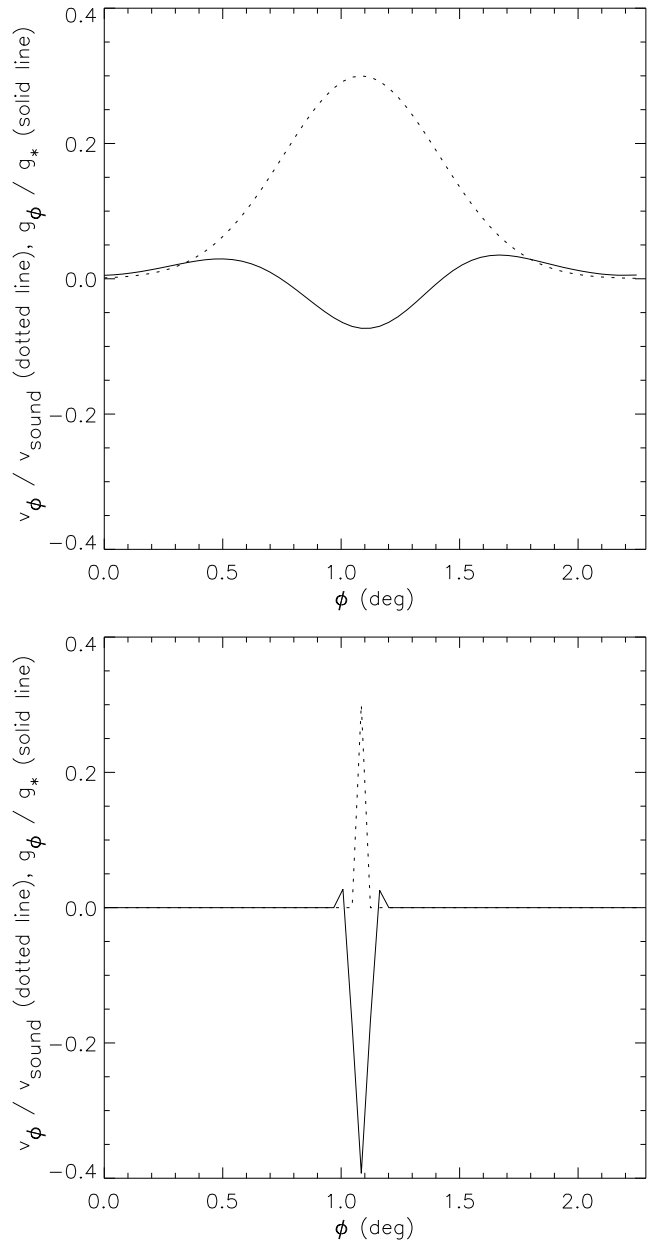}
\caption
{
Projected diffuse acceleration in the lateral direction from
the two lateral rays (solid line) for a long- (top) and
short- (bottom) wavelength velocity perturbation (dotted line).
For both cases, we use a relaxed CAK radial velocity.
}
\end{figure}

% \subsection{Numerical experimentations}
\subsection{3-Ray Nonlinear Simulation Results}

We have used this 3-ray method to carry out 2D simulations for the nonlinear
evolution of wind structure for four different combinations of the
two basic grid parameters, namely  the azimuthal resolution
$\delta \phi$ and the impact parameter $p$ of nonradial rays, with
values summarized in Table 1.
Both choices of the azimuthal grid size,
$\delta \phi = 1.33 \times 10^{-3}$ and $6.65 \times 10^{-4}$~rad,
are fine enough to well resolve the characteristic lateral Sobolev
angle, $l_{0}/ r\approx 0.01$.
The fine angle resolution is chosen in part because in this 3-ray
method, a smaller $\delta \phi$ implies, for any given $p$,
a finer radial resolution $\delta r$ (Eq. \ref{eq4}).
A larger $p$ also implies a finer radial grid (cf. Eq. \ref{eq4}),
as well as a more azimuthal orientation of the nonradial rays,
especially near the stellar surface;
this gives a greater sensitivity to lateral variations.
The initial investigations by Owocki (1999) found the relaxed wind
properties could depend quite strongly on the value of $p$.
Our two choices of $p=\sqrt{0.5}$ and $p=\sqrt{0.9}$ allow us to
further investigate this sensitivity.

In addition to the basic parameters, Table 1 also summarizes the
number of grid points needed to reach the fixed outer radius $R_{\rm
max} = 10 R_{\ast}$, as well as the extrema of the radial resolution
for each model.
The latter reveal that the ray-projected Sobolev length is very well
resolved at low heights, but unresolved near the outer grid radius.
Apart from these differing grid properties, all four models have the
same wind and stellar parameters as the models presented in DO-03,
and in the previous two sections.
To isolate physical effects that result solely from the use
of a 3-ray radiation force, we also run all four models using just
the radial ray transport;
these thus represent 2D-H+1D-R models, but run on the
(quite distinctive) 3-ray spatial grids.
% Finally, we find that the 3-ray models converge to a somewhat different mass loss rate
% as seen from the non-unity density contrast near the wind base in the left-column
% panels.
% This results from a slight normalization shift between the initial 1-ray CAK
% and the 3-ray non-Sobolev line-force.

Figure 7 shows a montage of grayscale images of the relaxed wind density
for the four models, each normalized to the starting conditions obtained
with an equivalent CAK model.
The top row shows the full radial extent
(to $R_{\rm max}= 10 R_{\ast}$),
while the bottom row zooms into the inner wind region
(to $3 R_{\ast}$).
The left (right) column corresponds to 3-ray (1-ray) simulations.
Each panel is divided into four wedges, corresponding respectively
to models A, B, C and D stacked clockwise from the vertical,
i.e.
in order of decreasing radial resolution.

For the lower resolution models (C and D), note that the shell structure
is never really disrupted, irrespective of the impact parameter $p$, and,
moreover, irrespective of the inclusion or neglect of lateral transport.
In contrast, for the higher resolution models (A and B), the role of the diffuse
force is quite apparent from a comparison of the left (3-ray) and right (1-ray)
columns, especially if one focuses on the bottom-row panels that
concentrate on the more well-resolved, inner wind.
Most notably, in this region the lateral coherence is much greater when
lateral transport is taken into account.
However, there does not seem to be a significant dependence on the choice of
impact parameter, with shell break-up starting at similar heights of ca.
2 $R_{\ast}$ in both models A and B.

We emphasize that, because of the degraded radial resolution and
increasingly radial orientation of the oblique rays, the structure forms
seen in the outer regions of these models are  probably not very realistic.
However, the inner regions have  high resolution in both angle
and radius, and thus we believe it is significant to find that
the models with lateral transport can retain a lateral coherence that
extends well beyond the azimuthal grid scale.
This is quite distinct from the results of both 2D-H+1D-R models
(cf. DO-03 Fig. 1 and the left and right columns of Fig. 7 here)
and the models of Sect. 2, which use a viscous form for the lateral
diffuse force.
These results thus encourage further investigation of the effect of lateral
transport in setting a finite lateral scale of instability-generated
flow structure in line-driven stellar winds.

\begin{figure*}[!ht]
 \vspace{18cm}
% \special{psfile=plot_3ray_4ble.eps}
% \special{psfile=2778f7.eps}
 \caption
{
Grayscale representation of the density contrast for relaxed non-Sobolev models
computed with the 3-ray (left column) and 1-ray (right column) approaches.
All models were run up until a total real time of 120,000 seconds.
Panels in the top (bottom) row show models only out to 10 (3) R$_{\ast}$.
Each panel is sub-divided into four wedges corresponding to models A, B C and D,
stacked clockwise from the vertical in order of decreasing radial resolution (Table. 1).
As before, each computed angular wedge is duplicated a number of times so that
the total angular extent of each wedge shown equals $22.5^{\rm o}$, i.e. a quarter
of the quadrant in each of the four panels.
The key result shown here is the presence of laterally coherent wind structures
close to the photosphere in the high resolution 3-ray models (A and B), while with the
equivalent 1-ray model, one observes instead wind structures with no noticeable lateral
coherence, i.e. of the order of the lateral numerical grid resolution.
}
\end{figure*}

\section{Conclusions, Summary, and Future Outlook}

This paper is a continuation of our ongoing efforts to develop
multi-dimensional models for the nonlinear evolution of structure
arising from the line-driven instability in hot-star winds.
Building upon the previous 2D-H+1D-R simulations of DO-03, which carry out 2D
hydrodynamics but use a 1D, radial form for the line-transfer and
force, we develop here approximate methods to account for two key
effects expected in multi-dimensional treatments of the line-force,
namely lateral line-drag and lateral averaging.

A summary of key results is as follows:

\begin{enumerate}

\item
The lateral drag effect of diffuse line-radiation can be well modeled
by casting the azimuthal line-force in a parallel viscosity
form that scales with the second spatial derivative of the azimuthal
velocity.
For viscous coefficients of order the values derived from linear
perturbation analysis, this does indeed lead to a strong damping of lateral
velocity variations in 2D, nonlinear simulations of the line-driven
instability.

\item
However, contrary to previous expectations, this reduction of lateral
flow does not lead to a coherence at a characteristic Sobolev angle
(ca. $1^{o}$) associated with the strongest damping.
Instead, it tends to further isolate the azimuthal zones of the model, and so
again leads to strong lateral incoherence down to the grid scale.

\item
Lateral coherence may instead arise from the lateral averaging
associated with the radial driving, as demonstrated here from
heuristic models that explicitly impart an azimuthal smoothing to
the radial driving force.

\item
To account more consistently for both the lateral line-drag and lateral
averaging effects, we have also experimented with a restricted,
grid-aligned, 3-ray method that provides an approximate nonlocal
treatment for both the radial and lateral components of the transport
and line-force.
Results for the highest resolution simulations indicate lateral averaging
does tend to sustain an extended lateral scale in the resulting flow
structure, but specifics of the results are left uncertain by the
inherent limitations of the method.

\end{enumerate}

Overall, a central conclusion is thus that lateral averaging of
diffusion radiation appears promising as an effect that could set a
distinct lateral scale to structures arising from the line-driven
instability.
However, further modeling with a less-restricted multi-ray method will be
needed to confirm this idea, and to determine quantitatively the likely scale.
In particular, as mentioned in the introduction and in DO-03,
Rayleigh-Taylor or thin-shell instability likely plays a competing role
in controlling such a lateral scale, although with a magnitude difficult to
assess at present.
We intend to explore development and application of such methods in
our continuing work within this overall effort, with results to be
reported in future papers in this series.

\acknowledgements
SPO acknowledges support of NSF grant AST-0097983, 
awarded to the University of Delaware.

\appendix

\section{Linear Analysis for Lateral Diffuse Force}

Some key insights into the role of the multidimensional
diffuse line force can be gleaned from the 3D linear perturbation
analysis carried out by ROC.
Let us begin with ROC's Eq. (A2) for the perturbed
radiative acceleration tensor with components
$T_{ij} \equiv \delta g_{i}/\delta v_{j}$,
giving the i'th component of the radiative acceleration $\delta g_{i}$
arising from the j'th component of the velocity perturbation
$\delta v_{j}$.
% which is assumed to have a sinusoidal variation with arbitrary vector
% wavenumber ${\bf k}= \{ k_{l} \}$.
For the diffuse component (stemming from the second
term in the square bracket),  we first note from symmetry and parity
arguments that the tensor is purely real and diagonal,
given by (for the pure scattering case $\epsilon=0$)
\beq
T_{ii} = - s \Omega  \left < n_{i}^{2}
{ ( n_{i} k/2 x_{\ast} Q_{0} )^{2} \over 1 + ( n_{i} k/2x_{\ast} Q_{\rm o})^{2} }
\right > \, ,
\label{A1}
\eeq§
where the angle brackets denote averaging over solid angle,
and we have assumed the usual case of longitudinal perturbations with
${\bf k}\,  || \, \delta {\bf v}$.
% and $C'$ absorbs multiplicative terms defined in ROC Eqs. (23) and (24).
The factor $Q_{0}$, defined in ROC Eq. (13), accounts for the angle
variations of the velocity gradient in the smooth, spherically
symmetric background flow, but for simplicity we have
approximated this here by its value $Q_{0} = v_{o}/(r v_{\rm th}) = 1/l_{0}$
at the isotropic expansion point, where
$\sigma \equiv d\ln v_{o}/d\ln r -1 = 0$.
This makes the order unity factor
$s = \left<D\right>/\left<D\mu\right> = 2/(1+\mu_{\ast})$,
% In this case,  the normalization constant takes the simple form,
% \beq
% C' ={ g_{o} \over v_{\rm th} } { 2 \over 1+\mu_{\ast} }
% % { 1 \over Q_{0} }
% \, ,
% \label{A2}
% \eeq
where $\mu_{\ast} \equiv \sqrt{1-R_{\ast}^{2}/r^{2}}$.
The overall factor $-s \Omega$ gives a net damping rate that scales
with the radial instability growth rate,
$\Omega \equiv 2 x_{\ast} \alpha g_{o}/v_{\rm th}$,
where $\alpha \approx 0.6$ is the CAK exponent.
For simplicity we henceforth take the blue-edge frequency
$x_{\ast}$ to be unity.

In this work, we are interested in the lateral components $T_{11}=T_{22}$,
representing variations along the azimuthal direction,
for which the direction cosine in standard spherical coordinates
($\theta$,$\phi$) takes the form
$n_{1} \equiv n_{\phi} = \sin \theta \cos \phi$.
Then defining $K\equiv k/2Q_{0}$, the integrals required
for evaluation of the angle averaging take the form,
\beq
{\delta g_{\phi} \over \delta v_{\phi} } = -
%  { g_{o} \over v_{\rm th} } { 2 \over 1+\mu_{\ast} }
{ s \Omega \over 4 \pi }
\int_{0}^{2 \pi} \, d\phi \,
\int_{-1}^{1} \, d\mu \,
{ K^{2}  \cos{^{4} \phi} \, (1-\mu^{2})^{2}
 \over 1 + K^{2} \cos{^{2} \phi} \, (1-\mu^{2}) }
\, ,
\label{A2}
\eeq
where $\mu \equiv \cos \theta$.
From numerical evaluation of these integrals, we find that this can be
approximated (within ca. 10\%) by the simple form,
\beq
{\delta g_{\phi} \over \delta v_{\phi} }
\approx  - { s \Omega \over 3 } \, { K^{2} \over 1 + K^{2} }
= -
% \left (
{ s \Omega \over 3 }
% \, { 2/3 \over 1+\mu_{\ast} } \right )
~
{k^2 \over 4 Q_{0}^{2} + k^{2}}
\, .
\label{A3}
\eeq

In the short-wavelength limit $k \gg 2 Q_{0}$, this gives
\beq
{\delta g_{\phi} \over \delta v_{\phi} }
\approx - \Omega_{damp}
\equiv - {s  \Omega \over 3 }
= - {2 \alpha s  g_{o} \over 3 v_{\rm th} }
% \, { 2/3 \over 1+\mu_{\ast} }
~~ ; ~~ k \gg 2 Q_{0}
\, ,
\label{A4}
\eeq
where $\Omega_{damp}$ represents a damping rate associated with the
lateral line-drag of this diffuse radiation.

In the long-wavelength limit $k \ll 2 Q_{0}$, we find
\beq
{\delta g_{\phi} \over \delta v_{\phi} }
\approx - { \Omega_{damp}  k^{2} \over 4Q_{0}^{2}}
\approx - (\alpha s/6) \, (v_{\rm th} r ) \, k^{2}
~ ; ~ k \ll 2 Q_{0}
\, ,
\label{A5}
\eeq
where the latter equality uses the approximation
$g_{o} \approx v_{o}^{2}/r$.
Note that this is just the form that would arise from a parallel viscosity with
kinematic viscosity coefficient proportional to $ v_{\rm th} r$, as
assumed in Eq. (\ref{eq1}).
% Note that this
% \beq
% g_{\phi} = r v_{th} {\partial^{2} v_{\phi} \over r^{2} \partial \phi^{2}}
% \label{A7}
% \eeq
This thus forms a key motivation for the lateral viscosity
approach used in Sect. 2 to model the 2D nonlinear development of
structure arising from the strong radial instability of line driving.

The limiting forms (\ref{A4}) and ({\ref{A5}) of Eq. (\ref{A3}) also
provide a means to understand the small- and large-scale linear
perturbation tests done for the 3-ray SSF method in Sect. 4.2.
% (as displayed in lower and upper panels of fig. 6).
For the case in which the Gaussian perturbation width $r \Delta \phi_{o} = 0.0005 \, r$
is much  smaller than the lateral Sobolev length
$l_{0} = r v_{\rm th}/v_{o} \approx 0.01 \, r$, the lower panel of Fig. 6
shows that $g_{\phi} \approx - 0.5 g_{\ast} v_{\phi}/v_{\rm th}$,
which for a characteristic CAK acceleration $g_{o} \approx g_{\ast}/(1-\alpha)$
is quite consistent with the scaling
given in Eq. (\ref{A4}) for this small-scale limit.

However, for the larger scale perturbation, the upper panel of Fig. 6
shows that the force response scales as the second derivative of the
original Gaussian velocity perturbation,
consistent with the viscous scaling predicted in the large-scale limit
of Eq. (\ref{A5}).

\end{document}